\documentclass[aps,prb,twocolumn,superscriptaddress,showpacs,floatfix,reprint]{revtex4-1}

\usepackage[dvips]{graphicx}
\usepackage{color}
\usepackage{latexsym}
\usepackage{amsmath}
\usepackage{amsthm}
\usepackage{amsfonts}
\usepackage{amssymb}
\usepackage{mathptmx}
\usepackage{subfigure}

\usepackage{mathrsfs}
\usepackage{natbib}

\providecommand*{\unit}[1]{\,\ifmmode \mathrm{\,#1}\else\textup{#1}\fi}

\begin{document}

\title{Quantum phase-slip junction under microwave irradiation}

%
%
%
%
%
%
\author{A. Di Marco}
\affiliation{Universit{\'e} Grenoble Alpes, LPMMC, F-38000 Grenoble, France}
\affiliation{CNRS, LPMMC, 25 Avenue des Martyrs B.P. 166, F-38042 Grenoble Cedex, France}

\author{F.\ W.\ J.\ Hekking}
\affiliation{Universit{\'e} Grenoble Alpes, LPMMC, F-38000 Grenoble, France}
\affiliation{CNRS, LPMMC, 25 Avenue des Martyrs B.P. 166, F-38042 Grenoble Cedex, France}

\author{G. Rastelli}
\affiliation{CNRS, LPMMC, 25 Avenue des Martyrs B.P. 166, F-38042 Grenoble Cedex, France}
\affiliation{Zukunftskolleg, Fachbereich Physik, Universit{\"a}t Konstanz, D-78457, Konstanz, Germany}
%
%
%
%
%
%
%
%
%

\begin{abstract}

We consider the dynamics of a quantum phase-slip junction (QPSJ) -- a dual Josephson junction -- connected to a microwave source with frequency $\omega_\textrm{mw}$. With respect to an ordinary Josephson junction, a QPSJ can sustain dual Shapiro steps, consisting of well-defined current plateaus at multiple integers of $ e  \omega_\textrm{mw} / \pi$ in the current-voltage ($I$-$V$) characteristic. The experimental observation of these plateaus has been elusive up to now. We argue that thermal as well as quantum fluctuations can smear the $I$-$V$ characteristic considerably. In order to understand these effects, we study a current-biased QPSJ under microwave irradiation and connected to an inductive and resistive environment. We find that the effect of the fluctuations is governed by the resistance of the environment and by the ratio of the phase-slip energy and the inductive energy. Our results are of interest for experiments aiming at the observation of dual Shapiro steps in QPSJ devices for the definition of a new quantum current standard.

\end{abstract}

%
%
%
%
%
%
%
%
%
%
%
%
%
%
%
%
%
%
%
%
%
%
%
%

\date{\today}

\maketitle

\section{Introduction}

The Josephson junction (JJ) is one of the most used superconducting devices in low-temperature condensed matter experiments. A single JJ is the building block of various sensors and electronic components\cite{Barone:1982,Likharev:1986,Tinkham:1996,SQUIDS_2:2006,Fulton:1989} and plays an important role in quantum computation and information.\cite{Makhlin:2001,Devoret:2004,You:2005,Martinis:2009,Ladd:2010} On a more fundamental level, JJs with small capacitance have become paradigmatic systems for studying decoherence and dissipation of a quantum particle coupled to the external world and for analyzing the transition from quantum to classical states.\cite{Schmid:1983,Caldeira:1983,Caldeira:1983erratum,Devoret:1985,Leggett:1985,Leggett:1987,Schon:1990,Leggett:2002,Schlosshauer:2010,Weiss:2012}

Many of the JJ applications are based on the Josephson effect: a Cooper-pair tunneling current $I_J$ can flow through a JJ in the absence of an applied bias voltage. The amplitude of this supercurrent is a non-linear function of the phase difference $\varphi$ between the two superconductors of the junction, $I_J=I_c\sin(\varphi)$. The critical current $I_c$ is the maximum Cooper-pair current that can be carried by the junction.
A voltage drop $V_J=(\hbar/2e) d\varphi / dt$ appears across the junction when $\varphi$ changes as a function of time.
The classical dynamics of $\varphi$ is ruled by the equations of motion for a fictitious particle moving in a tilted washboard potential. In particular, a phase-locking effect can occur when the JJ is irradiated with microwaves of frequency $\omega_\textrm{mw}$.\cite{Kautz:1996}
Then, the so-called Shapiro steps of constant voltage $V_{J,m}=m(\hbar/2e)\omega_\textrm{mw}$, with $m$ integer, appear in the current-voltage characteristic in addition to the zero-voltage supercurrent state.\cite{Shapiro:1963,Shapiro:1964} These steps are related only to the fundamental constants of physics (the Planck constant $\hbar$ and the electron charge $e$) and are used in metrology to define the quantum voltage standard.\cite{Taylor:1969,Hamilton:2000,Flowers:2004,Scherer:2012} The necessary metrological accuracy is reached at low temperatures and using junctions with large Josephson energy $E_J=\Phi_0 I_c/(2\pi) \sim 100\unit{meV}$ [$\Phi_0=h/(2e)$ is the superconducting flux quantum] and small charging energy $E_C=e^2/2C\sim 10\unit{neV}$, where the capacitance of the junction $C$ plays the role of the inertial mass in the dynamics of the phase. Moreover, the JJ is typically embedded in a circuit whose resistance $R \alt R_Q$, with $R_Q=h/(4 e^2)=6.45\unit{k\Omega}$ the superconducting resistance quantum. Under these conditions, thermal and quantum fluctuations of the phase $\varphi$ are suppressed efficiently.\cite{Kautz:1996,Kohlmann:2011}

The Josephson junction has an exact dual counterpart, the so-called quantum phase-slip junction
(QPSJ).\cite{Likharev:1985-1,Likharev:1985-2,Averin:1985,Averin:1990,Golubev:1992,Mooij:1999,Mooij:2005,Mooij:2006,Zazunov:2008,GuichardHekking:2010}
Physical realizations of QPSJ that have been discussed in the literature are a single Josephson junction with a finite
capacitance\cite{Likharev:1985-1,Likharev:1985-2,Averin:1985,Averin:1986,Averin:1990,Zazunov:2008,GuichardHekking:2010}
or a linear chain of such Josephson junctions,~\cite{GuichardHekking:2010,Matveev:2002,Pop:2010,Rastelli:2013,Weissl:2014}
and a narrow superconducting nanowire.\cite{Hriscu:2011_1,Hriscu:2011_2,Vanevic:2011,Astafiev:2012,Hongisto:2012,Webster:2012,Peltonen:2013}
With respect to an ordinary JJ, the role of the phase and the charge in a QPSJ is interchanged. Specifically, Cooper-pair tunneling is replaced by its dual process, i.e., the slippage by $2\pi$ of the phase difference between two well-defined superconducting regions of the device. As a consequence, the relations governing the behavior of a QPSJ are exactly dual to the usual Josephson relations.
The voltage $V_J=V_c\sin(\pi q/e)$ across the QPSJ is a non-linear function of the charge variable $q$, where the critical value $V_c$ is the maximum voltage that the junction can sustain.
The Cooper-pair current $I_J=dq/dt$ is different from zero only for time-dependent $q$. As a consequence, under microwave irradiation, a QPSJ should sustain a set of current steps, i.e., the dual Shapiro steps $I_{J,m}=m \, e \omega_\textrm{mw} / \pi$.~\cite{,Averin:1990,Golubev:1992}

Although the occurrence of coherent phase-slip events has been reported for small Josephson junctions irradiated with microwaves,\cite{Kuzmin:1991,Haviland:1991,Kuzmin:1994}
a clear experimental evidence of the dual steps has been elusive so far.
Indeed, the dual Josephson relations pertain to a QPSJ with a relatively well defined charge $q$, achieved when phase-slips are produced at an appreciable rate,
a condition which is not easily compatible with the existence of a well-defined underlying superconducting state.
Actual realizations of a QPSJ are typically operated in a regime where $V_c$ is not large, so that charge fluctuations are important, and may well mask the dual Shapiro steps.

In this paper, we study the role of both thermal and quantum fluctuations of charge on the properties of the dual Shapiro steps. We present the results of a combined analytical and numerical analysis of a QPSJ irradiated with microwaves and embedded in a resistive ($R$) and inductive ($L$) electromagnetic environment. We will see, in particular, that an important role is played by the inductance $L$, the quantity dual to the capacitance $C$ of a usual Josephson junction. By duality, we expect that the fluctuations of the charge $q$ are governed by the ratio $U_0/E_L$ of the characteristic phase-slip energy $U_0 =2eV_c/(2\pi)$, dual to the Josephson coupling energy $E_J$, and the inductive energy $E_L = \Phi_0^2/(2L)$, dual to the charging energy $E_C$ of a Josephson junction.~{\cite{Mooij:2006}
The larger $L$, the smaller $E_L$ and the larger the ratio $U_0/E_L$, thus favoring a well-defined charge state of the QPSJ. Recent experiments on nanowires~\cite{Astafiev:2012,Peltonen:2013} and chains of Josephson junctions~\cite{Weissl:2014} typically achieve $U_0/E_L$ ratios that are of the order of $10^{-2} - 10^{-1}$. We will analyze the microwave response of a QPSJ in this regime in detail and study in particular the resolution and accuracy of the dual Shapiro steps.

%
%
\begin{figure}[t]
\centering \subfigure[]
{\raisebox{7mm}
{\includegraphics[scale=0.4,angle=0.]{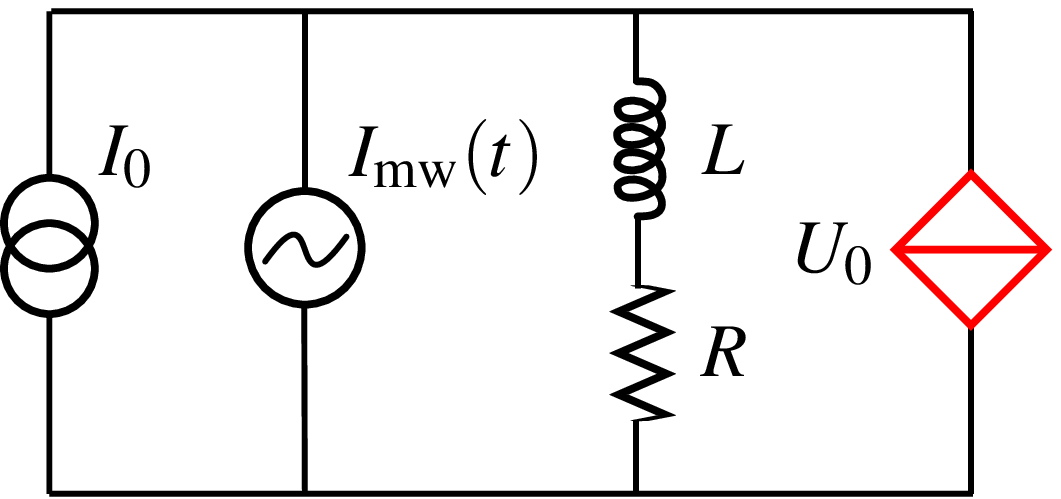}}
}
\hspace{2mm} \subfigure[]
{
\includegraphics[scale=0.35,angle=0.]{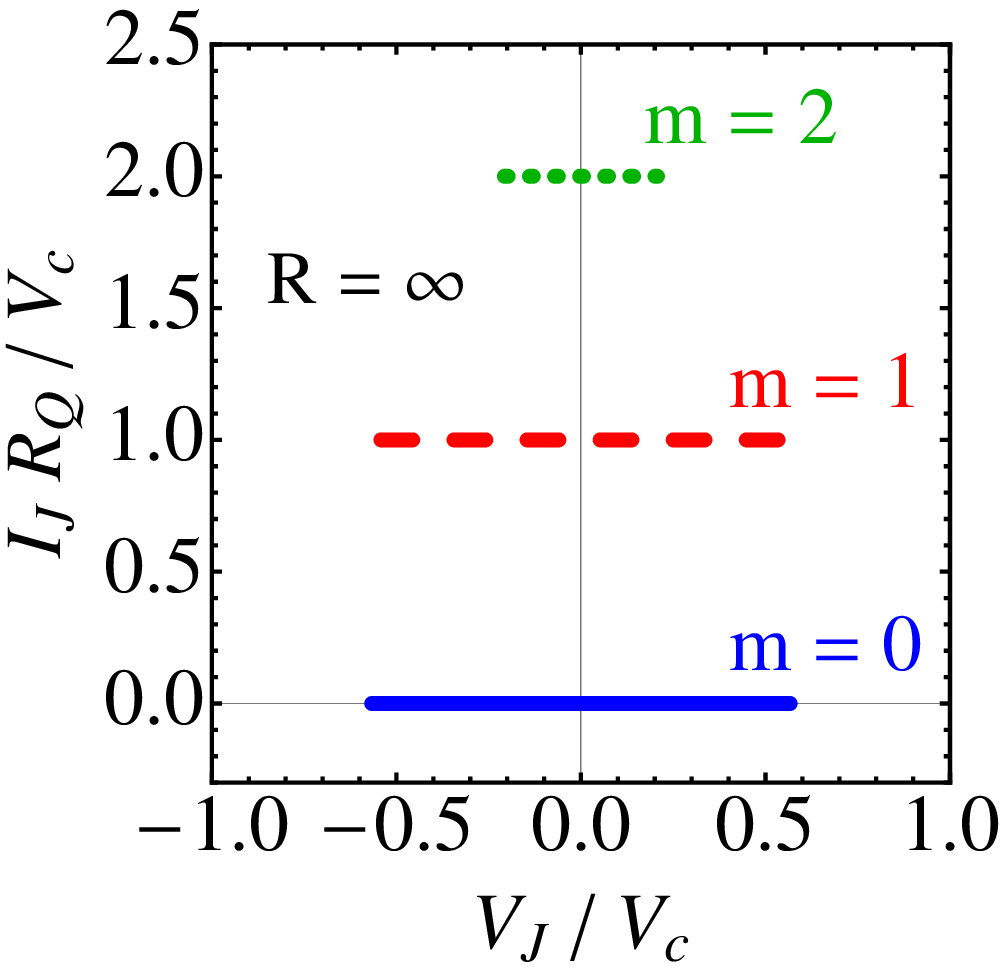}
}
\caption{(Color online)
(a) Circuit of a current-biased QPSJ with phase-slip energy $U_0 = e V_c/\pi$ embedded in a resistive ($R$) and inductive ($L$) electromagnetic environment. The total bias current is the sum of a DC component $I_0$ and an AC one $I_\textrm{mw} (t)$. (b) Dual Shapiro steps corresponding to $m=0$, 1, and 2 for a QPSJ in the absence of the environment $(R=\infty)$. Parameters: $\alpha=1.4$ and $\hbar \omega_\textrm{mw} /2 \pi U_0 = 1$.
}
\label{fig1}
\end{figure}
%
%
%

\section{Qualitative discussion of the main results}

We consider a current-biased QPSJ, with critical voltage $V_c=\pi U_0 /e$, which is connected in parallel to a resistive ($R$) and inductive ($L$) electromagnetic environment, and is driven by both a DC source, $I_0$, and an AC one $I_\textrm{mw} (t) = I_\textrm{mw} \cos \omega_\textrm{mw} t$ with amplitude $I_\textrm{mw}$ and microwave frequency $\omega_\textrm{mw}$ [see Fig.~\ref{fig1}(a)]. This circuit is related to the voltage-biased version by the Th{\'e}venin-Norton theorem setting $I_0=V_0/R$ and $|I_\textrm{mw}| = |V_\textrm{mw}|/\sqrt{R^2 + L^2 \omega_\textrm{mw}^2}$.
The results for the $I_J$-$V_J$ curve of the QPSJ of Fig.~\ref{fig1}(a) that will be discussed in the following
are independent of the specific choice of the external bias.

Let us first consider the case when the environment is absent, $R \to \infty$, in the circuit of Fig.~\ref{fig1}(a).
Then the dual Josephson relations describing the current-biased QPSJ can be straightforwardly integrated yielding
\begin{equation}
V_J^{(\textrm{mw})}(\omega_B) = V_c \sum \limits_{m=-\infty}^{+\infty} J_m(\alpha) \sin{( \pi q_0/e + \omega_B t + m \omega_\textrm{mw} t)},
\label{qualvjideal}
\end{equation}
where $J_m$ is a Bessel function of the first kind.
We defined the parameters $\alpha = \pi I_\textrm{mw}/(e\omega_\textrm{mw})$ and $\omega_B = \pi I_0/e$.
From this result we see that whenever the DC bias current $I_0$ equals $m e\omega_\textrm{mw}/\pi$,
the QPSJ will sustain a charge-dependent DC voltage $V_{J,m} = V_c J_m(\alpha) \sin (\pi q_0/e)$.
 In other words, phase-locking occurs, leading to the appearance of a dual Shapiro step in the DC current-voltage characteristic of the QPSJ
which is located at $I_{J,m} = m e\omega_\textrm{mw}/\pi$, and whose width is given by $2 V_c J_m(\alpha)$ [see Fig.~\ref{fig1}(b)].

When the resistance $R$ of the environment is restored, it is generally a challenging task to analyze the behavior of the QPSJ, due to the
interplay of quantum dissipative dynamics, non-linearity and a time-dependent signal.
The combined effect of the application of microwaves and the presence of charge fluctuations induced by the resistive-inductive environment
was discussed in literature in several limits. For instance, assuming the charge has a
classical dynamics, one can study a Langevin equation (see Appendix~\ref{app-A})
as discussed in the seminal work Ref.~\onlinecite{Likharev:1985-1}.
Quantum corrections were considered in Ref.~\onlinecite{Zazunov:2008} but in the absence of microwaves.
A full quantum approach of the dynamics of the junction in the presence of microwaves 
was discussed in Refs.~\onlinecite{Averin:1986,Golubev:1992} using a density matrix approach.
Thermal and quantum fluctuations play a crucial role as they significantly
affect the shape of the current-voltage characteristic.

%
%
\begin{figure}[tb]
{\includegraphics[scale=0.74]{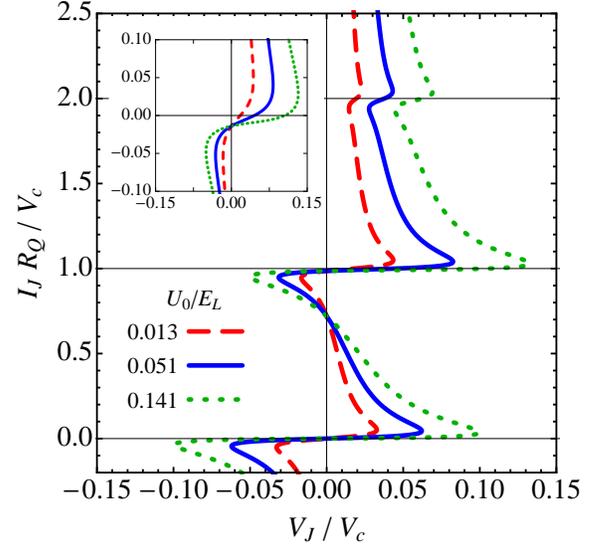}}
\caption{(Color online) $I_J$-$V_J$ characteristic obtained from the numerical evaluation of Eq.~(\ref{eq:First_Order}) in the low-conductive regime, $g=0.2$. Parameters: $k_BT/U_0 = 0.25$,  $\alpha=1.4$ and $\hbar\omega_\textrm{mw} / 2\pi U_0 =1$.
The three curves are obtained using three different values of the inductance of the environment such
that $U_0/E_L = 0.013$ (red dashed line), $U_0/E_L = 0.051$ (blue solid line), $U_0/E_L = 0.141$ (green dotted line).
The inset shows the relative deviation $\delta I_m = \pi I_J/me \omega_\mathrm{mw} -1$ for the first step, $m=1$.}
\label{fig2}
\end{figure}
%
%
%

We are now in a position to state the main results of this paper.
We use an approach that is non-perturbative in both the environmental coupling strength $g$, defined as $g= R_Q/R$, and the microwave coupling strength $\alpha$,
i.e., a generalization of the so-called $P(E)-$theory\cite{Ingold:1992} in the dual regime.~\cite{Averin:19902,Zazunov:2008}
As we will see in the following, this implies that analytical results can only be obtained in the limit $U_0/E_L < 1$. On the other hand, this corresponds to the relevant experimental situation where QPSJs are studied with relatively low phase-slip rates and not too large inductances. In the limit $U_0/E_L <1$, we find that, at the first order in $U_0$, the QPSJ's current-voltage characteristic in the presence of microwaves can be straightforwardly obtained from the DC result without microwaves (see Appendix~\ref{appb}):
\begin{equation}
\label{eq:First_Order_intro}
V_J^{\textrm{(mw)}}(\omega_B)
= \sum_{m=-\infty}^{+\infty} J_m^2 \left( \alpha \right) \, \,
V_J^{\textrm{(DC)}}  \left( \omega_B + m \omega_\textrm{mw} \right) \  ,
\end{equation}
in agreement with a general result recently demonstrated in Ref.~\onlinecite{Safi:2014}. Here $V_J^{\textrm{(DC)}}$ is given by Eqs.~(\ref{eq:First_Order_NoBias}) and (\ref{eq:Dual_PE}) and corresponds to the voltage as obtained by means of the $P(E)$-theory, as discussed in Appendix~\ref{appb}.
Specifically, this result implies that the current-voltage characteristic of a QPSJ with $U_0/E_L <1$ under microwave irradiation
is obtained by replicating the known DC characteristic of the QPSJ in the absence of microwaves at the positions of the current
plateaus $I_{J,m} = m e\omega_\textrm{mw}/\pi$, which are expected for a QPSJ in the absence of the environment.
Equation~(\ref{eq:First_Order_intro}) was derived previously in Ref.~\onlinecite{Golubev:1992} for the particular case where the QPSJ is a Josephson junction.
In this paper, we analyze the consequences of Eq.~(\ref{eq:First_Order_intro}) in detail. In particular,  we discuss 
the role of the inductance $L$ of the external impedance in view of the accuracy of the Shapiro steps and consider the effect of Joule heating produced in the resistive environment.

We focus on the case $ g < 1$, for which dual Shapiro steps clearly appear in the $I$-$V$ curve.
Figure \ref{fig2} displays typical current-voltage characteristics obtained in this situation, taking $g=0.2$.
We see that the current-voltage characteristics are strongly modified in the simultaneous presence of microwaves and charge fluctuations induced by the environment:  Rather than being a set of discrete steps, the QPSJ's $I_J$-$V_J$ characteristic is a continuous curve, connecting subsequent steps, bending back towards a zero-voltage state in between them.
In other words, in the presence of microwaves, a replica of the relevant QPSJ's DC current-voltage characteristic (see Appendix~\ref{appb} and Fig.~\ref{fig8}) is found for each dual Shapiro step.
As expected, in the presence of charge fluctuations, the width of the steps is smaller than the value $2 V_c J_n(\alpha)$, found for $g=0$;
also, the dual steps are no longer strictly horizontal but acquire a small but finite linear slope.
Note the role played by the inductance $L$, which limits the effects of the charge fluctuations.
As it is clearly seen in Fig.~\ref{fig2}, the larger $L$, the larger the width of the steps and the smaller their slopes.
This can be seen in particular in the inset of Fig.~\ref{fig2}, which presents the relative accuracy $\delta I_m \equiv (I_J - I_{J,m}) / I_{J,m} = \pi I_J/me \omega_\mathrm{mw} -1$ for the first Shapiro step, $m=1$.
The inset also shows that the accuracy of the dual step is not only limited by charge fluctuations but also by a systematic shift of the step position, down by about 0.0015 in relative accuracy.
This is due to the finite overlap of the various replicas.
The shift can be reduced by increasing the microwave frequency so that the replicas are more separated along the $I_J$-axis, thus reducing their overlap.

The rest of the paper is structured as follows. In Sec.~\ref{sec:JJdual}, we introduce the model Hamiltonian for a QPSJ connected to a microwave source. We also show the results of the perturbation theory for the dissipative coupling with the external environment and for the coupling with the applied microwaves. In Sec.~\ref{sec:Expansion}, we develop the non-perturbative approach. In Sec.~\ref{sec:first_order}, we discuss the results to the leading order in the $U_0$ expansion focusing on the accuracy of the dual Shapiro steps and on the Joule heating effects. We draw our conclusions in Sec.~\ref{sec:conclusions}.

\section{Current-biased QPSJ}
\label{sec:JJdual}

\subsection{QPSJ Hamiltonian}

The Hamiltonian of the current-biased QPSJ in the circuit depicted in Fig.~\ref{fig1}(a) is given by
\begin{equation}
\label{eq:H_dual}
\hat{H} \! = \! -U_0 \cos\left[ \frac{\pi}{e}  \left( \hat{q}  + \hat{Q}_{RL} \right) \right]
-  \frac{\hbar I(t)}{2e} \hat{\varphi} +
\hat{H}_\textrm{env} \left[ \{\hat{Q}_{\lambda}\}, \{ \hat{\varphi}_{\lambda} \}  \right]  \ .
\end{equation}
Here the charge and phase operators $\hat{q}$ and $\hat{\varphi}$ are canonically conjugate, satisfying the commutation relation $[\hat{\varphi},\hat{q}] = 2 e i$. As a consequence, $\hat{q}$  satisfies the equation of motion $\dot{\hat{q}} = I(t)$ and thus corresponds to the total charge injected into the parallel combination of the QPSJ and the $R$-$L$ environment. The first term in Eq.~(\ref{eq:H_dual}) describes the nonlinear QPSJ with phase-slip energy $U_0$,  which carries the charge $\hat{q} + \hat{Q}_\textrm{RL}$, where the charge variable $\hat{Q}_\textrm{RL}=\sum_\lambda \hat{Q}_\lambda$ accounts for the charge of the dissipative $R$-$L$ environment. We model it using an infinite ensemble of harmonic oscillators (Caldeira-Leggett model),~\cite{Caldeira:1983,Caldeira:1983erratum} described by the third term of Hamiltonian (\ref{eq:H_dual}),
\begin{equation}
\label{eq:H_env}
\hat{H}_\textrm{env}\left[ \{\hat{Q}_{\lambda}\}, \{ \hat{\varphi}_{\lambda}  \} \right]
=\sum_{\lambda=1}^{+\infty}\left[ \frac{\hat{Q}_\lambda^2}{2C_\lambda} + \frac{1}{2L_\lambda}\left( \frac{\hbar \hat{\varphi}_\lambda}{2e} \right)^{\!\!2} \right]
\, .
\end{equation}
The charge $\hat{Q}_\lambda$ and the phase $\hat{\varphi}_\lambda$ represent the momentum and position, respectively, of the
$\lambda$-oscillator with characteristic frequency $\omega_\lambda=1/\sqrt{L_\lambda C_\lambda}$. According to the fluctuation-dissipation theorem, $(1/2)\langle [\hat{I}_\textrm{RL}(t),\hat{I}_\textrm{RL}(0)]_+\rangle_\omega = \hbar \omega \Re\mbox{e} [Y(\omega] \coth(\hbar \omega \beta/2)$, where $\hat{I}_\textrm{RL} = \dot{\hat{Q}}_\textrm{RL}$ is the fluctuating current in the $R$-$L$ environment and $[...,...]_+$ denotes the anticommutator. This yields the relation
\begin{equation}
\label{eq:bath_parameters}
\Re\mbox{e}[Y(\omega)] = \pi \omega^2 \sum \limits_{\lambda} \sqrt{\frac{C_\lambda}{L_\lambda}} \delta(\omega^2 - \omega_\lambda^2) \ ,
\end{equation}
linking the parameters of the Caldeira-Leggett bath with the environmental admittance of the circuit in Fig.~\ref{fig1}(a):
\begin{equation}
Y(\omega) = 1/(R - i\omega L) \, .
\label{adm}
\end{equation}
Finally, the coupling between the charge operator $\hat{q}$ and the bias current $I(t)$ is given by the second term in (\ref{eq:H_dual}).

Hamiltonian (\ref{eq:H_dual}) has been used to describe QPSJs based on nanowires,~\cite{Mooij:2006} Josephson junctions~\cite{Zazunov:2008} and chains
of Josephson junctions.~\cite{GuichardHekking:2010}
In Appendix~\ref{appa}, we show how Eq.~(\ref{eq:H_dual}) can be obtained starting from the well-known Hamiltonian of a current-biased single Josephson junction connected to a R-L impedance.

\subsection{Current-Voltage Characteristic}

The DC current $I_J$ flowing through the QPSJ element is given by the difference between the total DC current $I_0$ and the current flowing through the $R$-$L$ impedance of the circuit of Fig.~\ref{fig1}(a),
\begin{equation}
\label{eq:I-V}
I_J = I_0 - V_J/R   \ .
\end{equation}
Here, $V_J$ is the DC component of the voltage drop across the QPSJ element. Using the Josephson relation between $\hat{\varphi}$ and $V_J$ and the Heisenberg equation of motion for the operator $\hat{\varphi}$ generated by the Hamiltonian $\hat{H}$, this potential reads as
\begin{equation}
\label{eq:Voltage}
V_J = \frac{\hbar}{2e}  {\left< \frac{d\hat{\varphi}}{dt} \right>}_\textrm{DC}
= V_c  {\left< \sin \left[ \frac{\pi }{e } \left(  \hat{q} + \hat{Q}_{RL} \right) \right] \right>}_\textrm{DC} \  .
\end{equation}
The symbol $\left< \dots \right>$ denotes the quantum statistical average for the system described by the Hamiltonian $\hat{H}$, Eq.(\ref{eq:H_dual}).

\subsubsection{Dual Shapiro steps in the absence of environment}
\label{subsec:ideal}

By setting $\hat{Q}_{RL}=0$ in Eq.~(\ref{eq:H_dual}), the coupling with the environment vanishes and the system
corresponds to an ideal current-biased QPSJ whose Hamiltonian $\hat{H}_0$ contains only the first two terms of $\hat{H}$.
Introducing a complete set of discrete phase-states for the QPSJ, $\left| \phi_n \right> = 2 \pi \left| n \right>$ with $n$ integer,  we can express $\hat{H}_0$ as
\begin{equation}
\label{eq:H_exact}
\hat{H}_{0} =
-\frac{U_0}{2}
\sum_n
\left(
\left| n \right> \left<  n+1 \right| + \textrm{H.c.}
\right)
- \frac{\hbar  I(t)}{2e}
\sum_n 2\pi n  \left| n \right> \left<  n \right|  \  ,
\end{equation}
in the phase representation. When $I_{ \textrm{mw}}=0$, Eq.~(\ref{eq:H_exact}) corresponds to the well-known Wannier-Stark ladder problem for a particle moving in a tilted tight-binding lattice (see Fig.~\ref{fig3}). The tilt $I_0$ provides an energy difference equal to $\hbar \omega_B$ between two adjacent phase states. The term proportional to $U_0$ induces transitions between adjacent phase states, i.e., phase-slip events. In the absence of microwaves or a coupling to the environment, we have only coherent Bloch oscillations and the associated energy difference $\hbar \omega_B$ can not be accommodated by the system. Hence no finite DC component is found for the voltage $V_J$ in this case.

Switching on the microwave field, the tilted lattice acquires an additional, oscillatory slope with amplitude  $I_\textrm{mw} \neq 0$. For this problem, the unitary evolution operator can be evaluated exactly and it reads as~\cite{Gluck:2002,Korsch:2003}
\begin{equation}
\label{eq:U_exact}
\hat{U}(t) =
e^{i \mathcal{Q}(t) \hat{n} }
e^{
i \frac{U_0}{2\hbar} \int^{t}_0 dt' \left[ \hat{K} \exp\left(i \mathcal{Q}(t')  \right) + \hat{K}^{\dagger} \exp\left(-i\mathcal{Q}(t')  \right)  \right]
} \, ,
\end{equation}
in which we set
\begin{equation}
\label{eq:Q_t}
\mathcal{Q}(t) = \omega_B \, t + \alpha \sin\left( \omega_{ \textrm{mw}}  t  \right)  \, .
\end{equation}
In Eq.~(\ref{eq:U_exact}), we also introduced the number operator $\hat{n} = \sum_n n \left| n \right>\left< n \right| $ and the ladder operator $\hat{K} = \sum_n \left| n \right>\left< n + 1\right| $.
After some algebra, the expectation value of the voltage operator in Eq.~(\ref{eq:Voltage}) on the state $\hat{U}(t)\left| q_0 \right>$, the time evolved initial quasi-charge state $|q_0\rangle$, is
\begin{align}
\label{eq:perfect_dual_shapiro}
V_J^{(\textrm{mw})}(t) & =  V_c \ \sin\left[ \pi q_0/e +  \mathcal{Q}(t) \! \right] \nonumber \\
& =  V_c \sum_{m=-\infty}^{+\infty} J_m\left(  \alpha \right) \sin\left(  \frac{\pi q_0}{e}  +  \omega_Bt   +  m \omega_\textrm{mw} t  \right) \ .
\end{align}
Equation (\ref{eq:perfect_dual_shapiro}) coincides with Eq.~(\ref{qualvjideal}) and describes the ideal dual Shapiro steps: a non-vanishing DC-voltage now appears
each time the bias-current $I_0=I_J$ satisfies the condition $I_J = m e \omega_\textrm{mw} / \pi$, as shown in Fig.~\ref{fig1}(b).
The dual Shapiro steps, labeled with the index $m=0,\pm 1, \dots$, are replicas of the zero-voltage state obtained with $m=0$ and $\alpha=0$, rescaled with the corresponding Bessel function of the first kind $J_m(\alpha)$. The coherent emission/absorption of microwave photons with energy $\hbar \omega_{ \textrm{mw}}$ is at the origin of this phenomenon.
The local phase states undergo a coherent quantum tunneling upon exchanging the energy $\hbar \omega_\textrm{mw}$ with the microwave field [see Fig.~\ref{fig3}(a)].

%
%
\begin{figure}[tb]
\centering \subfigure[]
{\includegraphics[scale=0.27]{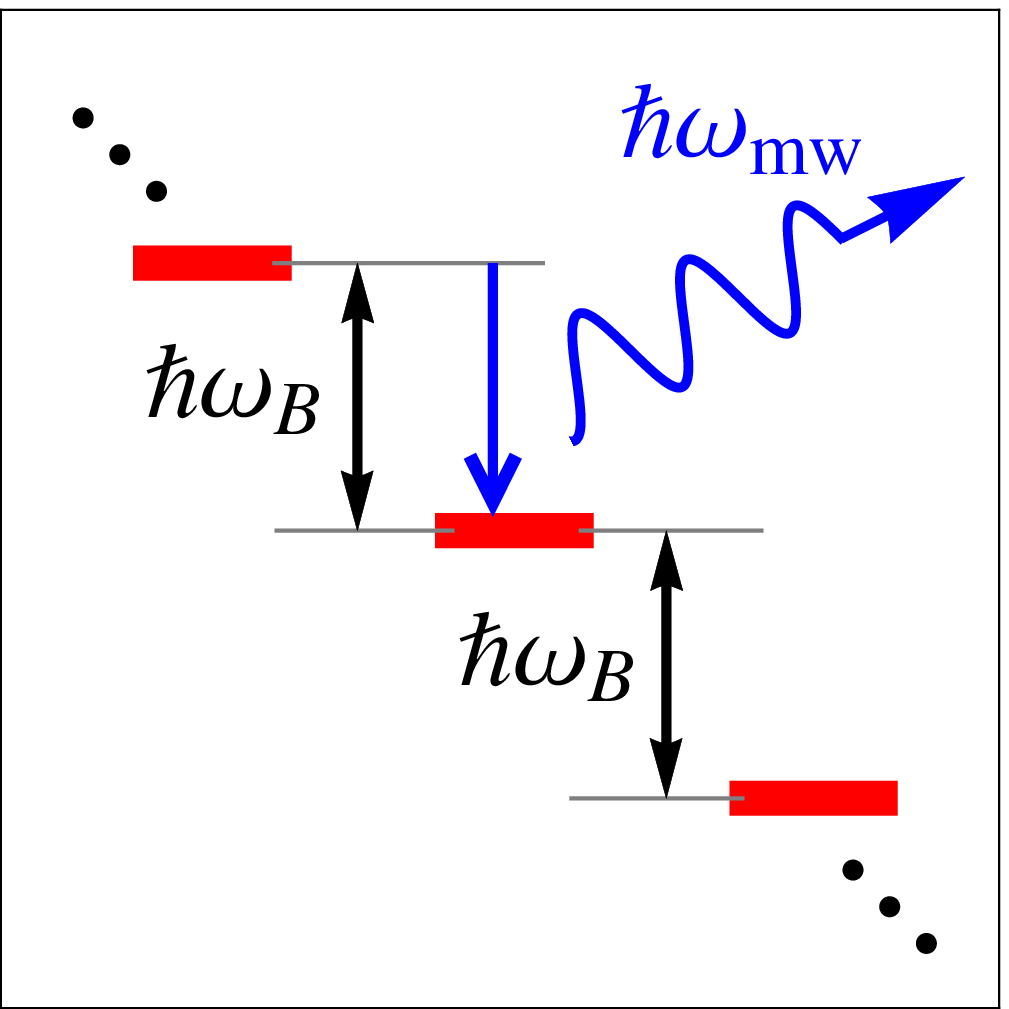}}
\hspace{1mm}\subfigure[]
{\includegraphics[scale=0.27]{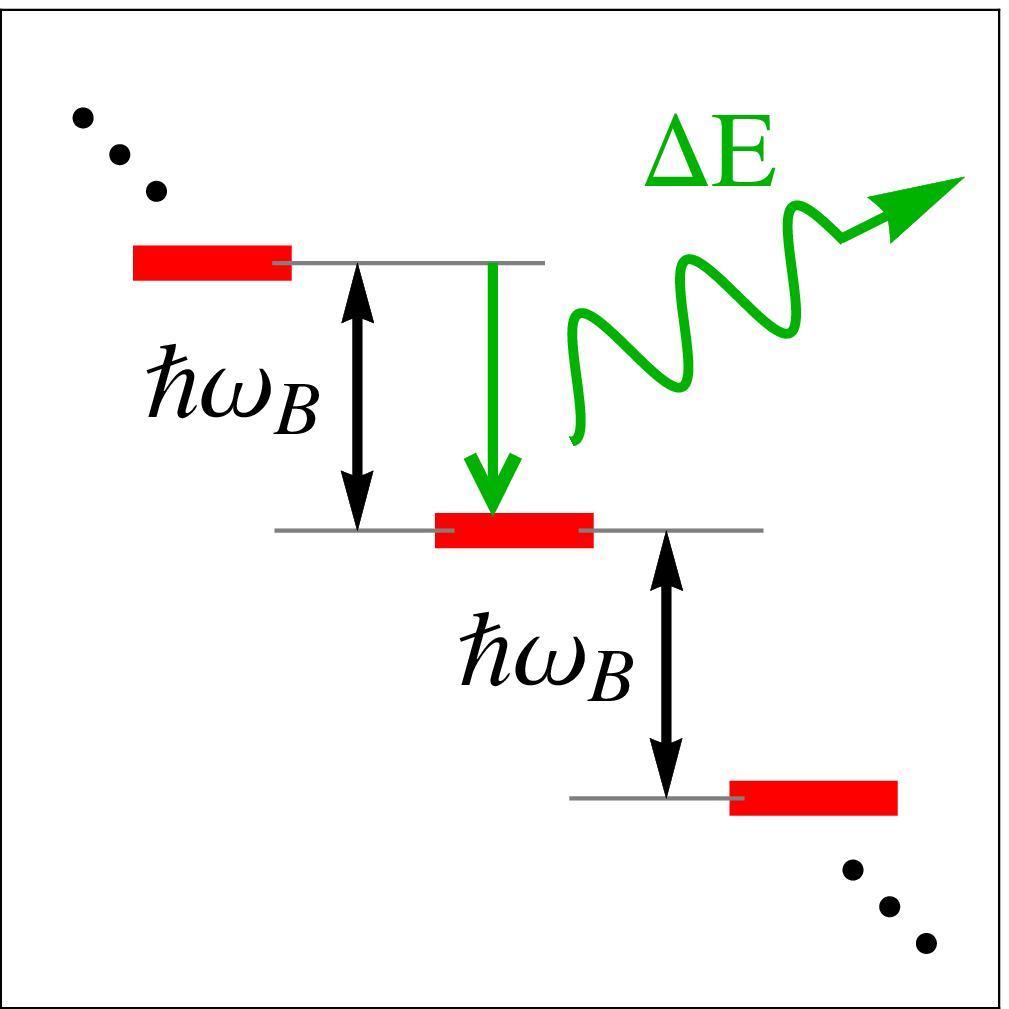}}
\hspace{1mm}\subfigure[]
{\includegraphics[scale=0.27]{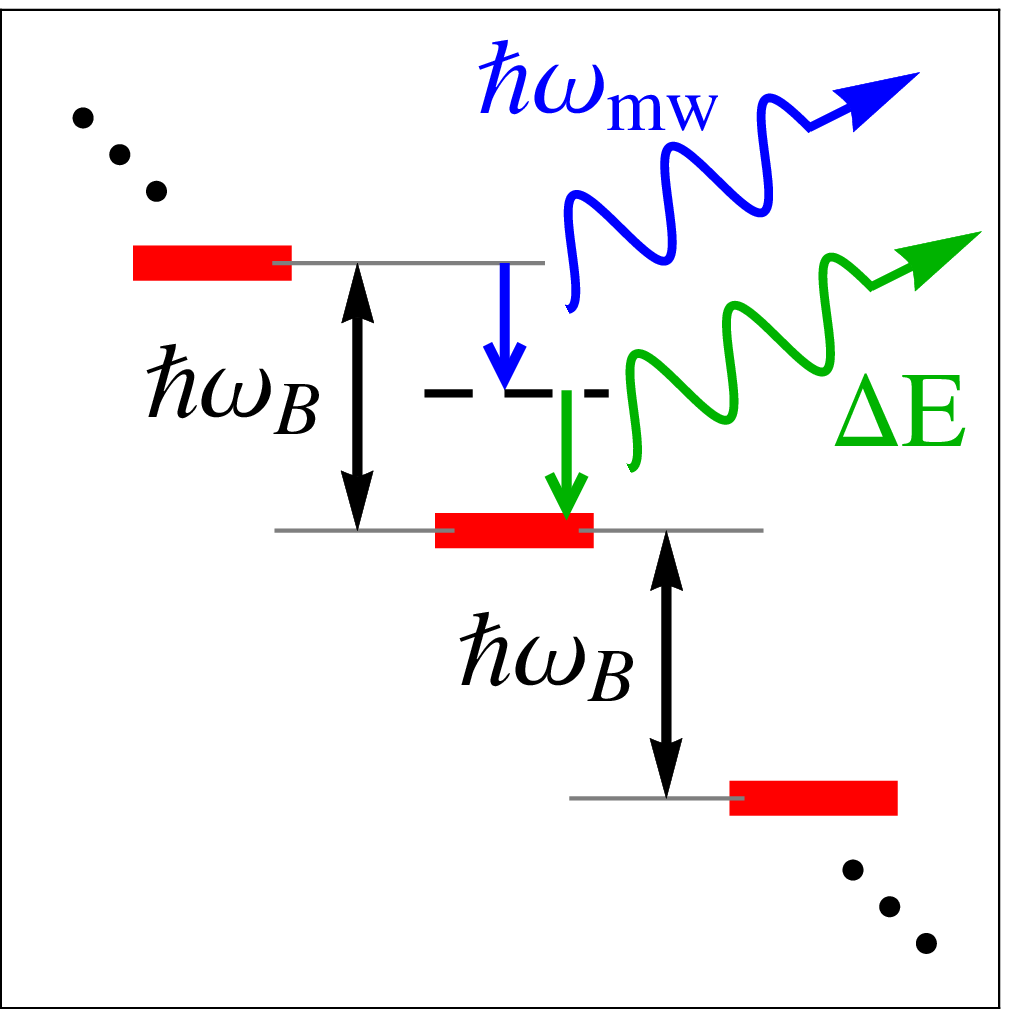}}
\caption{(Color online)
Wannier-Stark ladder. The tilt provided by the bias current $I_0$ induces an energy separation $\hbar \omega_B$ between adjacent
phase states indicated by red horizontal bars.
(a) Phase-locking occurs when the resonant condition $\omega_B = m \omega_\textrm{mw}$ is satisfied.
For $m=1$, a photon with energy $\hbar\omega_\textrm{mw}$ is exchanged with the microwave source.
(b) Environment-assisted transitions between adjacent states in the Wannier-Stark ladder lead also to the appearance of a finite voltage
across the QPSJ element. (c) Wannier-Stark ladder in the presence of both microwave and environmental photons with
energies $\hbar \omega_\textrm{mw}$ and $\Delta E$ respectively.}
\label{fig3}
\end{figure}
%
%
%

\subsubsection{Perturbation theory}

We next analyze the current-voltage characteristic of the QPSJ in terms of perturbation theory in microwave interaction $\alpha$
and dissipative coupling $g$. We show that this approach systematically leads to divergent behavior.
For simplicity, we assume the bath to be at zero temperature.

Applying the unitary transformation $\hat{U}_\textrm{env}=\exp\left[- i\hat{\varphi}\hat{Q}_\textrm{RL}/2e  \right]$ to Hamiltonian (\ref{eq:H_dual}), we obtain the QPSJ Hamiltonian in the form $\hat{H}_s^{'} = \hat{H}_0 + \hat{H}_{int}$ in which we consider as the unperturbed Hamiltonian
\begin{equation}
\hat{H}_0= -U_0 \cos\left( \frac{\pi}{e} \hat{q} \right) - \frac{\hbar I_0}{2e} \hat{\varphi} \, ,
\end{equation}
and the interaction term
\begin{equation}
\hat{H}_{int}= - \frac{\hbar I_\textrm{mw}}{2e} \cos(\omega_\textrm{mw} t) \hat{\varphi}  +
\hat{H}_\textrm{env} \left[ \{\hat{Q}_{\lambda}\}, \{ {\hat{\varphi}_{\lambda} + \hat{\varphi}} \}  \right] \, .
\label{H_int_un}
\end{equation}
In this canonical form, the voltage across the QPSJ is given by
\begin{equation}
\label{eq:Voltage_simple}
V_J  = V_c  {\left< \sin \left( \frac{\pi }{e }  \hat{q}  \right) \right>}_\textrm{DC}  \, .
\end{equation}
Using the interaction picture,  we expand the unitary time evolution operator in terms of $\hat{H}_{int}$ [Eq.~(\ref{H_int_un})] to calculate $V_J$ [Eq.~(\ref{eq:Voltage_simple})].
After some algebra, for vanishing microwave strength $\alpha=0$, we obtain for the DC component of the voltage
\begin{equation}
\label{eq:divergence_1}
V_J^{(\textrm{DC})} =  g V_c^2/(2R_Q I_0) \, .
\end{equation}
This result is  linear in $g$ and corresponds to the first order expansion of the solution of the classical
Langevin equation for the charge [see Eq.~(\ref{classsol}) in Appendix~\ref{app-A}].
Its validity requires $V_J/V_c \ll 1$, hence $I_0 \gg g V_c/R_Q$.
We conclude that perturbation theory breaks down in the limit of vanishing DC current bias.

%
%

In the presence of microwaves, $\alpha \ne 0$, the result (\ref{eq:divergence_1}) generalizes to
\begin{equation}
\label{eq:divergence_2}
V_J^{(\textrm{mw})} =  \frac{g V_c^2}{2R_Q} \sum_{m=-\infty}^{+\infty}   \frac{J_m^{2}(\alpha)}{I_0 +  m e \omega_\textrm{mw} / \pi} \, ,
\end{equation}
which shows that the divergent behavior found for $I_0\to 0$ is repeated at the positions $I_0 \to m e \omega_\textrm{mw}/\pi$
at which the dual Shapiro steps are expected.

Although the perturbative approach is divergent and is inappropriate to describe the dual Shapiro steps in the presence of dissipation, it is useful for giving a simple picture of the QPSJ's dynamics: The incoherent tunneling of the localized phase states in the Wannier-Stark ladder generally occurs via the combined emission and/or absorption of a certain number of photons with energy $\hbar \omega_\textrm{mw}$ of the microwave source and the exchange of an amount of energy $\Delta E$ with the thermal bath (see Fig.~\ref{fig3}).
One expects that the interplay between the photon-assisted and environment-assisted phase-slippage causes the smearing of the ideal dual
Shapiro steps.  Indeed, the sharp resonance condition $ \hbar\omega_B = \hbar\omega_\textrm{mw}$
associated to the single microwave photon emission can not be fulfilled anymore as the QPS junction can now dissipate
the energy $\hbar\omega_B$ at any bias current because the energy difference $ \Delta E = \hbar(\omega_B - \omega_\textrm{mw})$
is emitted in the environment (see Fig.~\ref{fig3}).

\section{Dual Shapiro's steps: non-perturbative approach}
\label{sec:Expansion}

We now develop a theory to describe the combined effect of charge fluctuations induced by the environment on one hand and microwave irradiation on the other hand, which is non-perturbative in both $g$ and $\alpha$.
To determine the current-voltage characteristic of the QPSJ by means of Eq.~(\ref{eq:I-V}), we need the DC component $V_J$ of the potential across the QPSJ given by Eq.~(\ref{eq:Voltage}).
The quantum statistical average in the right-hand side of Eq.~(\ref{eq:Voltage}) can be calculated by means of the Keldysh formalism.\cite{Zwerger:1987,Grabert:1998,Ingold:1999,Zazunov:2008}
Introducing the so-called Keldysh closed-time contour $C_k$ which goes from $t=-\infty$ to $t_0$ and back to $t=-\infty$ and treating the cosine term in Eq.~(\ref{eq:H_dual}) as a perturbation, one can obtain an exact series expansion in terms of the QPSJ energy $U_0$.
In this expansion, the contribution of the oscillators forming the harmonic bath decouples from the contribution of the QPSJ charge $q$ so that one can evaluate the quantum statistical averages exactly to each order.
We have generalized this solution taking into account the presence of the microwave signal.
The time-dependent voltage across the QPSJ reads as
\begin{equation}
\label{eq:Expansion}
\frac{V_J(t_0)}{V_c}
\!\!
=
\!\!
\sum_{n=0}^{+\infty}
\!
\frac{(-1)^n}{2i}
\!
\left( \frac{U_0}{\hbar} \right)^{2n+1}
\!\!\!\!\!\!
\sum_{\{ \eta_k\}}
\!
\int_{-\infty}^{t_0}\!\!\!\!\! dt_1
\!
\dots
\!
\int_{-\infty}^{t_{2n}}\!\!\!\!\!dt_{2n+1}
F_{\textrm{env}}  \ F_{\textrm{q}}\ ,
\end{equation}
where the term $F_{\textrm{env}}$,
\begin{equation}
\label{eq:F_env}
F_{\textrm{env}} =
e^{\sum_{k=1}^{2n+1}\sum_{k'=0}^{k-1}\eta_k \eta_{k'} M\left( t_{k'}-t_k \right)}
\prod_{k=1}^{2n+1}
\!\!
\sin\left[\sum_{k'=0}^{k-1} \eta_{k'} A( t_{k'}-t_k )\right] \ ,
\end{equation}
accounts for the environment-assisted phase-slip events and $F_\textrm{q}$,
\begin{equation}
\label{eq:F_bias}
F_\textrm{q}
=
e^{i \sum_{k=0}^{2n+1} \mathcal{Q}\left(\eta_k t_k\right)  }
=
\exp\left\{i \sum_{k=0}^{2n+1}\left[\omega_B \eta_k t_k  +  \alpha \sin\left( \omega_\textrm{mw} \eta_k t_k \right)  \right] \right\}  \  ,
\end{equation}
is related only to the free dynamics of the charge $q$ as given by Eq.~(\ref{eq:Q_t}). The dichotomic variables $\eta_k=\pm 1$, with $k=0,1,\dots,2n+1$, satisfy the constraint $\sum_{k=0}^{2n+1}\eta_k=0$ and the sum $\sum_{\{\eta_k \}}$ over all the possible configurations of $\eta_k$ stands for the product of the $2n+2$ sums $\sum_{\eta_1=\pm} \cdot\cdot\cdot \sum_{\eta_{2n+1}=\pm}$.

The functions of time $M(t)$ and $A(t)$ in Eq.~(\ref{eq:F_env}) describe the exchange of energy between the QPSJ and the external electromagnetic environment.
They determine
\begin{equation}
\label{eq:J_0}
J(t)=-M(t)-i \ \textrm{sign}(t)A(t) \, ,
\end{equation}
 i.e., the charge-charge correlation function
\begin{equation}
\label{eq:J_1}
J(t) = \sum_\lambda \left<  \hat{Q}_\lambda (t)\hat{Q}_\lambda (0) - \hat{Q}_\lambda^2 (0)  \right> \ ,
\end{equation}
which quantifies the fluctuations of the tunneling phase due to the thermal bath.\cite{Ingold:1992,Ingold:1994} In particular, $J(t)$ gives the coupling strength between the QPSJ and the environment. For the current-biased configuration of Fig.~\ref{fig1}(a), we have
\begin{equation}
\label{eq:J}
J(t)
=
2 R_Q \int_{-\infty}^{+\infty}    \frac{d\omega}{\omega} \Re\mbox{e}[Y(\omega)]
\left( \frac{e^{-i\omega t}-1}{1-e^{-\hbar\omega\beta}} \right)
 \, ,
\end{equation}
with $\beta=1/k_B T$ the inverse temperature. An exact calculation yields\cite{Zazunov:2008,Grabert:1998}
\begin{equation}
\label{eq:A_RL}
A(t) = \pi g \left( 1 - e^{-\omega_\textrm{RL} |t|} \right)  \ ,
\end{equation}
\begin{equation}
\label{eq:M_RL}
M(t) =
2g\frac{\pi |t|}{\hbar\beta}
-A(t) \cot\left( \frac{ \hbar\omega_\textrm{RL} \beta }{2} \right)
+ 2g\sum_{n=1}^{+\infty} \frac{1}{n} \frac{1-e^{- \nu_n |t| }}{1-(\nu_n/\omega_\textrm{RL})^2}  \ .
\end{equation}
Here $\nu_n = 2\pi n /\hbar\beta$ is the $n$-th Matsubara frequency, and $\omega_\textrm{RL}=R/L$ is the frequency scale characterizing the environment fluctuations at vanishing temperature.

The Jacobi-Anger expansion $\exp[i\alpha \sin(x)]=\sum_{m=-\infty}^{+\infty} J_m(\alpha)\exp[imx]$ allows to cast $F_\textrm{q}$
in terms of the Bessel functions of the first kind $J_m(\alpha)$,
\begin{equation}
\label{eq:Bessel_1}
F_\textrm{q}
\! =\!\!\!\!\!\!\!
\sum_{m_0=-\infty}^{+\infty}
\!\!\!\!\!\!\!\!
J_{m_0}(\alpha)
\! \dots \!\!\!\!\!\!\!\!\!\!
\sum_{m_{2n+1}=-\infty}^{+\infty}
\!\!\!\!\!\!\!\!\!\!
J_{m_{2n+1}}(\alpha)
\exp\!\!\left[i  \! \sum_{k=0}^{2n+1}  \!\! \left( \omega_B + \omega_\textrm{mw} m_k   \right)\eta_k t_k \right] \!.
\end{equation}
Performing the change of variables $\tau_k=t_{k-1}-t_k$, each time $t_k$ can be expressed as $t_k=t_0-\sum_{h=1}^{k}\tau_{h}$ with $k\geq 1$.
Then Eq.~(\ref{eq:Bessel_1}) becomes
\begin{eqnarray}
\label{eq:Bessel_2}
F_\textrm{q} &=& \sum_{\left\{ m_k \right\}} \left(\prod_{m_k}J_{m_k}\right) \,
\exp\left( i \omega_\textrm{mw} t_0  \sum_{k=0}^{2n+1} \eta_k m_k \right) \nonumber \\
&\times & \exp\left[-i  \sum_{k=0}^{2n+1}   \left( \omega_B + \omega_\textrm{mw} m_k   \right)\eta_k  \sum_{h=1}^{k} \tau_{h} \right]
\end{eqnarray}
where we used the sum rule $\sum_k \eta_k =0$.
Unlike the functions $M(t_{k'}-t_k)$ and $A(t_{k'}-t_k)$ in Eq.~(\ref{eq:F_env}) which depend only on
the time difference $t_{k'}-t_k=\sum_{h=1}^k\tau_h - \sum_{h'=1}^{k'}\tau_{h'}$, Eq.~(\ref{eq:Bessel_2}) is a function of the time $t_0$ at which we calculate the voltage across the QPSJ. From Eq.~(\ref{eq:Bessel_2}) we observe that the frequency spectrum of Eq.~(\ref{eq:Expansion}) at the time $t_0$
involves integer components of the single fundamental  frequency $\omega_\textrm{mw}$ applied to the dual junction.
This frequency mixing is due to the QPSJ which operates as a non-linear capacitance, i.e., it is related to the cosine dependence of the
QPSJ energy as a function of the charge $q$.
Thus, in the steady state regime, we can extract the DC component  by considering the time average of the general signal as
$ \overline{f(t)} = (1/T_{\textrm{mw}})  \int^{t_i+T_{\textrm{mw}}}_{t_i}  dt  \,\, f(t) $
 over a microwave period $T_{\textrm{mw}}=2\pi/\omega_{\textrm{mw}}$ where $t_i$ is an arbitrary initial time.
Then, the DC voltage reads as

\begin{equation}
\label{eq:Expansion_TimeAverage}
\frac{V_J}{V_c}
=
\frac{\overline{V(t_0)}}{V_c}
=
\dots \overline{F_{\textrm{q}}(t_0)}
=
\dots
\frac{1}{T_\textrm{mw}}\int_{t_i}^{t_i+T_\textrm{mw}} \!\!\!\!\!\!\!\!\!\!\!\!\! dt_0 \,\,
e^{i \omega_\textrm{mw} t_0  \sum_{k=0}^{2n+1} \eta_k m_k}  \ .
\end{equation}
The latter quantity is different from zero only if the sum rule
$
\sum_{k=0}^{2n+1} \eta_k m_k =0
$
is satisfied for each arbitrary configuration of the variables $\{ \eta_k\}$ at given set of the integers  $\{ m_k \}$
associated to the expansion of the Bessel functions.

\section{Lowest order results}
\label{sec:first_order}

A general analysis of the $U_0$-expansion Eq.~(\ref{eq:Expansion_TimeAverage}) is only possible in limiting cases. We focus here on the experimentally most relevant limit of relatively small QPSJ energy $U_0$, typically encountered in Josephson junction-based QPSJs. Then Eq.~(\ref{eq:Expansion_TimeAverage}) can be approximated with its first term. We discuss the range of validity of this approximation in the following. 

\subsection{Microwave irradiated QPSJ}

Considering $n=0$ only, the non-zero dichotomic variables are $\eta_0=\pm$ and $\eta_1=\pm$. Since they have to satisfy the constraint $\sum_k \eta_k = \eta_0+\eta_1=0$, it follows that the allowed configurations $\{ \eta_k \}=(\eta_0,\eta_1)$ are $(-,+)$ and $(+,-)$, i.e., $\eta_0$ and $\eta_1$ have opposite sign. This means that the time-average given by Eq.~(\ref{eq:Expansion_TimeAverage}) is different from zero if the indices $m_0$ and $m_1$ of the two possible sums of Bessel functions in Eq.~(\ref{eq:Bessel_1}) are equal. Then Eq.~(\ref{eq:Expansion_TimeAverage}) can be written as
\begin{equation}
\label{eq:First_Order}
V_J^\textrm{(mw)} (\omega_B)= \sum_{m=-\infty}^{+\infty} J_m^2 \left( \alpha \right) V_J^{(\textrm{DC})} \left( \omega_B + m \omega_\textrm{mw} \right) \  ,
\end{equation}
where $V_J^{(\textrm{DC})}$ is the voltage across the QPSJ in the absence of the microwaves [see Eq.~(\ref{eq:First_Order_NoBias}) of Appendix~\ref{appb}].
Thus, under the effect of the microwave radiation, the first-order voltage across the QPS junction is the superposition of an infinite number of zero-microwave potentials shifted by an integer multiple $m$ of $\omega_\textrm{mw}$. Unlike Eq.~(\ref{eq:perfect_dual_shapiro}), the weight of the $m$-th term in Eq.~(\ref{eq:First_Order}) is determined by the squared first-kind Bessel function of the $m$-th order, $J_m^2(\alpha)$.
This result is in agreement with the general theorem proved in Ref.~\onlinecite{Safi:2014} and was previously reported in Ref.~\onlinecite{Golubev:1992} for the particular case of a Josephson junction.
Since the sum rule $\sum_{-\infty}^{+\infty} J_m^2 \left( \alpha \right)=1$ holds,
the larger is $\alpha$, the smaller is the amplitude of the voltage corresponding to
$m=0$ and consequently the more important is the contribution of the higher-order terms.
In other words, changing the amplitude $\alpha$, the constant total weight re-distributes among the infinite terms of Eq.~(\ref{eq:First_Order}).

Using Eq.~(\ref{eq:I-V}) in combination with Eq.~(\ref{eq:First_Order}), we find that the $I_J$-$V_J$ characteristic of the QPSJ consists of $(m\omega_\textrm{mw})$-shifted and rescaled copies of the QPSJ's characteristic in the absence of microwaves [Eq.~(\ref{eq:First_Order_NoBias})] obtained for $I_\textrm{mw}=0$. These features occurring at $I_{J,m}=m \, e \omega_\textrm{mw} / \pi$ represent the dual Shapiro steps smeared by quantum and thermal fluctuations induced by the thermal bath. These results are shown in Fig.~\ref{fig2}, obtained by direct numerical evaluation of Eq.~(\ref{eq:First_Order_NoBias}) in combination with Eq.~(\ref{eq:First_Order}) for $g < 1$. The plotted smeared $I_J$-$V_J$ curves result from the competition and interference between the environment-assisted phase slippage and the pure photon-assisted tunneling of the phase induced by the microwave field. In order for these features to be resolved, the microwave frequency $\omega_\textrm{mw}$ has to be much larger than $\omega_B^\textrm{max} \approx  2\pi g / (\hbar \beta)$, the bias current corresponding to the back-bending point $(V_J^{(\textrm{DC})}/V_c)_\textrm{max}$ [see Eq.~(\ref{eq:Vj_max})].

%
%
\begin{figure}[th]
{\includegraphics[scale=0.75]{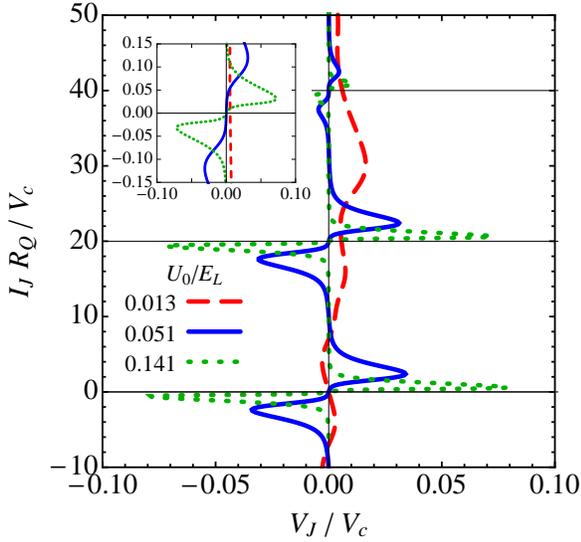}}
\caption{(Color online) $I_J$-$V_J$ characteristics obtained from the numerical evaluation of Eq.~(\ref{eq:First_Order}) in the high-conductive regime, $g=4.2$. Parameters: $k_BT/U_0 = 0.25$, $\alpha=1.4$, and $\hbar\omega_\textrm{mw} / 2\pi U_0= 20$.
The three curves are obtained using three different values of the inductance such that $U_0/E_L = 0.013$ (red dashed line), $U_0/E_L = 0.051$ (blue solid line), $U_0/E_L = 0.141$ (green dotted line). The inset shows the relative deviation $\delta I_m = \pi I_J/me \omega_\mathrm{mw} -1$ of the structure found for $m=1$ with respect to a perfect, first step obtained for $m=1$.}
\label{fig4}
\end{figure}
%
%
%

When $g>1$, the current-voltage characteristics of the microwave-irradiated QPSJ typically look like the ones plotted in Fig.~\ref{fig4}.
We find that they consist of replicas of the smeared current-voltage characteristics for $g>1$ and $I_\textrm{mw}=0$ (see Fig.~\ref{fig8}) centered around the positions of the ideal dual Shapiro steps.
Since the $I_J$-$V_J$ characteristics for $g>1$ are more smeared than the ones found in the low-conductive case, a higher microwave frequency
$\hbar \omega_\textrm{mw} /2 \pi U_0 =20$ has been used to resolve the various replicas in Fig.~\ref{fig4}.
When increasing the inductance $L$ for $g>1$, the smearing effects are reduced.
The inset of Fig.~\ref{fig4} shows the relative accuracy $\delta I_m = \pi I_J/me \omega_\mathrm{mw} -1$ of the structure found at $m=1$
when compared to a perfect, dual  step.
We see that the high conductance case does not produce single dual steps, but rather a doublet of two steps, located symmetrically around the value $me\omega_\mathrm{mw}/\pi$.
Combining Eq.~(\ref{eq:First_Order}) and the asymptotic result (\ref{eq:gaussian}), we expect the positions of the steps of the doublets to approach their asymptotic values $me\omega_\mathrm{mw}/\pi \pm \Phi_0/2L$ with increasing conductance $g$.
Eventually, a single dual  step is recovered for $L \rightarrow \infty$.

\subsection{Accuracy of the current  steps}
\label{subsec:accuracy}

%
%
\begin{figure}[t]
\centering \subfigure[]
{
   \includegraphics[scale=0.7]{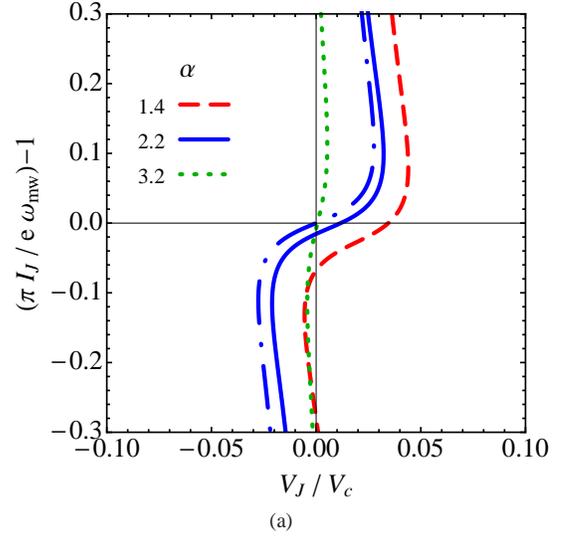}
}
\hspace{2mm} \subfigure[]
{
\includegraphics[scale=0.7]{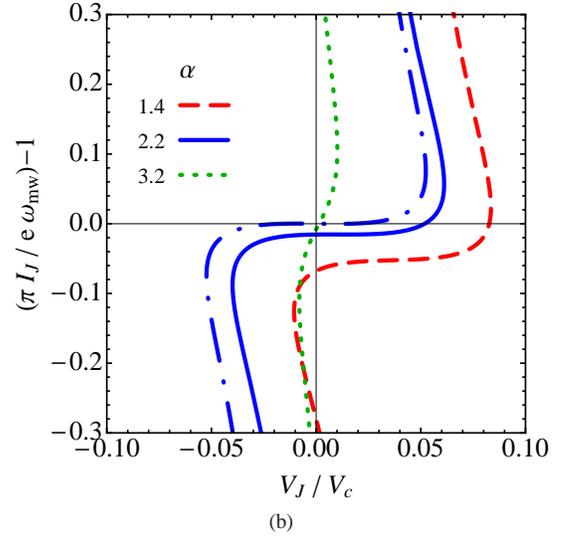}
 }
\caption{
(Color online) Relative deviation $\delta I_m = \pi I_J/me \omega_\mathrm{mw} -1$ for the first current step, $m=1$, for $k_B T / U_0 = 0.1$ and $\hbar \omega_\textrm{mw}/2 \pi U_0 = 0.16$. The ratio $U_0/E_L$ is equal to $0.0032$ for the plotted curves in panel (a) and $0.013$ for the ones in (b). In both panels, the $I$-$V$ characteristics have been obtained using three values of the microwave strength $\alpha$: $1.4$ (red dashed line), $2.2$ (blue solid and dotted-dashed lines) and $3.2$ (green dotted line). The (blue) dotted-dashed line corresponds to the unperturbed current step for $\alpha=2.2$ (see text).
}
\label{fig5}
\end{figure}
%
%
%

The reduction of quantum and thermal fluctuations affecting the current steps is crucial for their experimental observation as well as their potential applications,
such as in metrology.
In this respect, it is important to analyze the accuracy of the dual steps.
We focus on the relevant regime of low conductance, $g < 1$, where actual well-defined dual Shapiro steps are found and examine the smearing of the $m$-th step by considering the relative deviation $\delta I_m \equiv (I_J - I_{J,m}) / I_{J,m} =  \pi I_J/me \omega_\mathrm{mw} -1$.
Based on the asymptotic results of Eqs.~(\ref{eq:Slope_1}) and (\ref{eq:Vj_max}), we expect a minimal smearing when $T$ and $g$ are chosen as small as possible and $L$ large.

The behavior of $\delta I_m$ as a function of some of the relevant system parameters is studied numerically in Figs.~\ref{fig5}  and \ref{fig6} for the first dual
Shapiro step, $m=1$. In these figures, the solid, dashed and dotted lines correspond to three different microwave strengths $\alpha=$ 1.4, 2.2, and 3.2. Also shown (dashed-dotted line) is the behavior of the unperturbed dual Shapiro step for $\alpha=$ 2.2, i.e., $J_1^2(2.2) \times V_J^{(\textrm{DC})}(\omega_B-\omega_\textrm{mw})$, obtained by subtracting the contributions from all the other steps corresponding to $m \ne 1$ from the signal.

One sees that two phenomena generally limit the accuracy of the steps: (i) they are smeared around the actual plateau value and (ii) their position is offset with respect to the expected one. The latter phenomenon is absent for the unperturbed step:  indeed the shift of the step position is due to the finite overlap of the $m=1$  replica of the  Bloch nose with all the other replicas $m \ne 1$.
This suggests that increasing the microwave frequency should yield a better accuracy of the step position as it separates the replicas more, thereby reducing their overlap and, at the same time, improving their individual resolution. The result of an increasing of $\omega_\textrm{mw}$ on the step position can be seen by comparing Fig.~\ref{fig5}(b) with Fig.~\ref{fig6}.
We notice, for instance, that when $\alpha= 2.2$ the relative offset reduced from about 0.02 in the former to about 0.0004 in the latter by increasing $\omega_\mathrm{mw}$ by a factor of 10.

It is interesting to investigate why the curve for $\alpha = 2.2$ is less affected by the offset than the one for $\alpha = 1.4$, although the step size is the same for both curves. Indeed, the value of the squared Bessel functions $J_1^2(\alpha)$ determining the $m=1$ step width is almost equal for the two curves.
However the value $J_0^2(\alpha)$ is very different: $J_0^2(2.2) \approx 0.01$ whereas $J_0^2(1.4) \approx 0.32$.
In other words, the $m=0$ dual Shapiro step will strongly influence the step $m=1$ for $\alpha = 1.4$, leading to a large offset, whereas it
influences the $m=1$ step much less for $\alpha = 2.2$. The step corresponding to $\alpha = 3.2$ is more or less structureless, as its weight is very small, $J_1^2(3.2) \approx 0.07$.

As far as the smearing is concerned around the actual plateau position, a comparison between  Figs.~\ref{fig5}(a) and \ref{fig5}(b)
shows the effect of the inductance.
Increasing the inductance by a factor of 4 reduces the relative width of the step from about 0.1 in Fig.~\ref{fig5}(a) to about 0.05 in Fig.~\ref{fig5}(b).

%
%
\begin{figure}[t]
\includegraphics[scale=0.7]{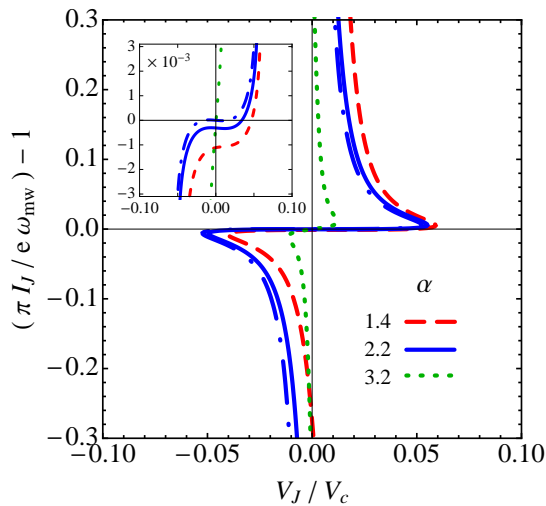}
\caption{
(Color online) Relative deviation $\delta I_m = \pi I_J/me \omega_\mathrm{mw} -1$ for the first Shapiro step, $m=1$, for $k_B T / U_0 = 0.1$,  $\hbar \omega_\mathrm{mw}/2 \pi U_0 = 2$ and $U_0/E_L = 0.013$. The plotted $I$-$V$ characteristics have been obtained using three values of the microwave strength $\alpha$: $1.4$ (red dashed line), $2.2$ (blue solid and dotted-dashed lines) and $3.2$ (green dotted line). The (blue) dotted-dashed line corresponds to the unperturbed Shapiro step for $\alpha=2.2$ (see text). The inset shows a close view of the steps plotted in the main panel.
}
\label{fig6}
\end{figure}
%
%
%

\subsection{The effect of Joule heating}
\label{subsec:heating}

%
%
\begin{figure}[th]
\centering \subfigure[]
{
\includegraphics[scale=0.7]{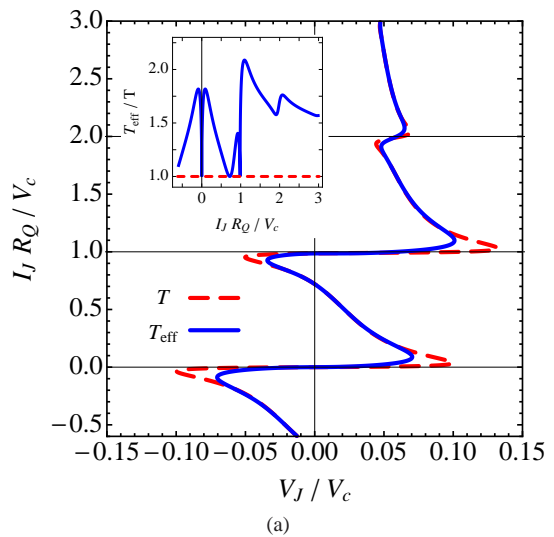}
}
\hspace{2mm} \subfigure[]
{
\includegraphics[scale=0.7]{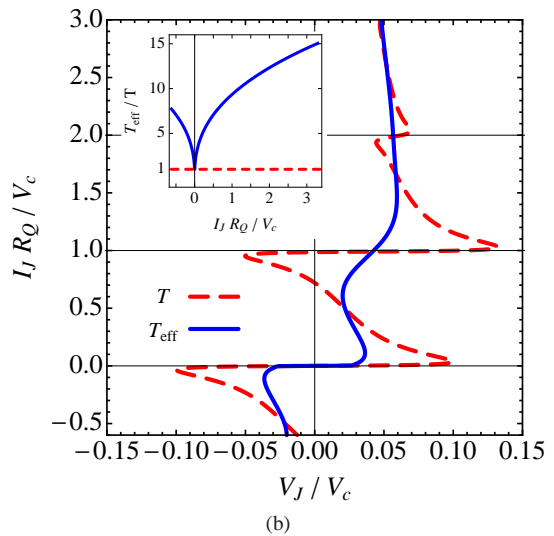}
 }
\caption{(Color online) Effect of Joule heating on the dual Shapiro steps obtained from the numerical evaluation of Eq.~(\ref{eq:First_Order}) in the low-conductive regime, $g=0.2$ for (a) the current-bias circuit Fig.~\ref{fig1}(a) and for (b) the equivalent voltage-bias circuit.
In both panels, for the (red) dashed $I_J$-$V_J$ curves the temperature is fixed to $k_B T / U_0 = 0.25$. The (blue) solid curves in (a) and (b) have been determined using the effective temperatures $T_\textrm{eff}$ which are the solutions of Eqs.~(\ref{eq:Teff}) and (\ref{eq:Teff_Vbias}) respectively, with $k_B T_\textrm{ph} / U_0 = 0.25$, and $U_0 = 4\unit{GHz}$. The electron-phonon coupling constant and the volume of the resistance $R$ are $\Sigma=10^9\unit{W m^{-3} K^{-5}}$ and $\Omega=10^{-19}\unit{m^3}$ respectively.~\cite{Webster:2012} All the $I$-$V$ characteristics in (a) and (b) are determined setting $U_0 / E_L =0.141$, $\hbar\omega_\textrm{mw}/2\pi U_0 =1$, and $\alpha = \pi I_\textrm{mw} / (e \omega_\textrm{mw}) = \pi V_\textrm{mw} / (e \omega_\textrm{mw} \sqrt{R^2 + L^2 \omega_\textrm{mw}^2}) = 1.4$,  as for the green dotted line in Fig.~\ref{fig2}. The insets show the rescaled effective temperature $T_\textrm{eff}/T$ as a function of the current through the QPSJ.}
\label{fig7}
\end{figure}
%
%
%

In this Section, we discuss an important aspect related to the experiment aimed to detect dual Shapiro steps, namely the effect of Joule heating in the $I$-$V$ characteristic of the QPSJ.~\cite{Astafiev:2012,Peltonen:2013,Weissl:2014,Webster:2012}

As we have seen above, we expect to approach the ideal dual Shapiro steps of Fig.~\ref{fig1}(b) under the condition $g\ll 1$. This means that the QPSJ is ideally embedded in a highly-dissipative environment. Such an environment is expected to produce also unwanted Joule heating which in turn would enhance the smearing of the steps. Indeed, in the low-conductance limit, $R\gg R_Q$, quantum effects due to the external bath become small, whereas thermal ones induced by heating may become dominant. In this context, the effective electronic temperature $T_\textrm{eff}$ of the $R$-$L$ series can be much larger than the phonon temperature $T_\textrm{ph}$. For the circuit of Fig.~\ref{fig1}(a), the current flowing through the $R$-$L$ branch is $V_J/R$, then the power dissipated by the resistance is $P_I=V_J^2/R$, where $V_J$ is a function of the temperature [see Eq.~(\ref{eq:First_Order})]. It follows that the effective temperature $T_\textrm{eff}$ can be estimated by the self-consistent equation~\cite{Wellstood:1994}
\begin{equation}
\label{eq:Teff}
T_\textrm{eff}^5=T_\textrm{ph}^5+ V_J^{ \ 2} (T_\textrm{eff}, \omega_B) / (R \ \Sigma \Omega) \ .
\end{equation}
In this last relation, $\Sigma$ is the material-dependent electron-phonon coupling constant, and $\Omega$ the volume of $R$. Figure~\ref{fig7}(a) shows the $I$-$V$ curve of a QPSJ embedded in an environment with $g\ll 1$ and fixed temperature, $k_B T / U_0 = 0.25$, where the Joule heating is not taken into account, together with the dual Shapiro steps smeared by the voltage-dependent effective temperature Eq.~(\ref{eq:Teff}) which accounts for the exchange of energy between the electrons and the phonons in the resistance $R$. We notice a reduction of the width of the steps, as one expects. From the inset of Fig.~\ref{fig7}(a), we see that $T_\textrm{eff}$ follows the oscillating trend of $V_J$. In particular, it coincides with $T_\textrm{ph}$ whenever $V_J=0$ and reaches its relative maxima for the values of $V_J$ around the maximum amplitude of the steps: the wider are the steps in the absence of Joule heating the larger is their effective thermal smearing.

On the other hand, Joule heating affects differently the dual Shapiro steps appearing in the $I$-$V$ characteristic of a voltage-biased QPSJ. In this configuration, the power $P_V=I_J^{ \ 2}R$, which is dissipated by the resistance $R$, is determined by the current flowing through both $R$ and the QPSJ, i.e.,
\begin{equation}
\label{eq:current_Vbias}
I_J=(V_0 - V_J) / R  \  .
\end{equation}
Here $V_J=V_J(T,V_0)$ is obtained from Eq.~(\ref{eq:First_Order}) replacing $I_0$ with $V_0/R$ and $|I_\textrm{mw}|$ with $|V_\textrm{mw}|/\sqrt{R^2 + L^2 \omega_\textrm{mw}^2}$. As a result, the effective temperature of the environment can be written as
\begin{equation}
\label{eq:Teff_Vbias}
T_\textrm{eff}^5 = T_\textrm{ph}^5+ I_J^{ \ 2} (T_\textrm{eff}, V_0) R / (\Sigma \Omega) \  .
\end{equation}
Inserting into Eq.~(\ref{eq:current_Vbias}) the temperatures $T_\textrm{eff}$ obtained by solving self-consistently Eq.~(\ref{eq:Teff_Vbias}) for different values of the DC voltage bias $V_0$, we obtain the (blue) solid QPSJ's $I$-$V$ characteristic shown in Fig.~\ref{fig7}(b). Notice that this curve is more smeared than the one found in the current-biased case and plotted in Fig.~\ref{fig7}(a) using the same set of parameters. As shown in the inset of Fig.~\ref{fig7}(b), $T_\textrm{eff}$ increases with $| I_J |$ and is equal to  $T_\textrm{ph}$ only when $I_J=0$. In particular, the effective temperature given by Eq.~(\ref{eq:Teff_Vbias}) is much larger than $T_\textrm{ph}$ when $I_J$ is close to $e \omega_\textrm{mw} / \pi$. Consequently, the Joule heating affects the steps for $m \neq 0$ more than the one occurring for $m=0$, as one can see from Fig.~\ref{fig7}(b), thereby compromising their experimental observation. The reduction of this effect is possible, for instance, with the decreasing of the microwave frequency $\omega_\textrm{mw}$. However, the use of smaller $\omega_\textrm{mw}$ leads also to the increasing of the offset of the steps which we discussed in Sec.~\ref{subsec:accuracy}.

In principle, Joule effect can be reduced by increasing the inductance $L$ of the environment rather than the resistance $R$. $L$ plays the same role of $R$ in the reduction of the fluctuations, as shown previously. As the dual Shapiro steps are replicas of the $I$-$V$ characteristic at low current, we can estimate the leading dependence for the smearing by considering Eq.~(\ref{eq:Slope_1}). We obtain the slope
$ G_0 R_Q \approx  2g {(k_BT/U_0)}^{2}  {(  E_L/U_0 )}^{2g }$ for $g \ll 1$.
We observe that the smearing due to the temperature can partially be compensated by increasing the inductance of the environment.

\section{Conclusions}
\label{sec:conclusions}

In this paper, we discussed the microwave response of a QPSJ embedded in an inductive-resistive environment. We focused on the regime of relatively small ratio of phase-slip energy $U_0$ over inductive energy $E_L$. The response consists of a series of well-defined current Shapiro steps, located at multiples of $e \omega_\mathrm{mw}/\pi$, if the environmental resistance is sufficiently large, such that the dimensionless conductance $g <1$. These steps are in fact replicas of the QPSJ's Bloch nose, observed in the absence of microwaves. Charge fluctuations induced by the environment smear the steps. This smearing can be reduced by decreasing the dimensionless environmental conductance $g$, decreasing the dimensionless temperature $k_B T/U_0$ and increasing the ratio $U_0/E_L$, which can be achieved by increasing the environmental inductance $L$. Finally, we showed that the conductance $g$ can not be decreased indefinitely, as heating effects may develop in the environment.

The results presented in this paper are relevant for recent experiments on Josephson junction chains~\cite{Weissl:2014} and nanowires.~\cite{Astafiev:2012,Peltonen:2013} In these works, typical phase-slip energies $U_0$ are in the range of $1 - 10 \unit{GHz}$, whereas the environmental inductances $L$ are $50 - 500 \unit{nH}$. This motivated the parameter choices used in this paper:
$U_0/E_L$ ranges from $0.001 -  0.1$; at typical cryostat temperatures $k_BT/U_0 \sim 0.1 - 0.2$. We found that, although dual Shapiro-type features could be visible experimentally for these parameters, their relative accuracy remains limited to about 0.001 by fluctuation effects.

To date, a systematic evidence for the existence of dual Shapiro steps is still lacking. The reason for this might well be that fluctuation effects have so far masked the steps for QPSJs with intermediate ratios of the parameter $U_0/E_L$ and not too small conductance $g$. Work on nanowire-based QPSJs with larger values of the ratio $U_0/E_L$ and lower conductances $g$ seems promising.~\cite{Hongisto:2012,Lehtinen:2012} At the same time these systems suffer from substantial heating effects.~\cite{Webster:2012} We conclude that further work is necessary, both on nanowires and on Josephson junction chains.

\acknowledgments

The authors thank W. Belzig, L. Glazman, W. Guichard, Yu. Nazarov, I. Safi, A. Zorin and especially C. Schenke for useful discussions. Financial support from the Marie Curie Initial Training Network (ITN) Q-NET (Project No. 264034), the European Research council (Grant No. 306731), Institut universitaire de France, and the EU FP7 Marie Curie Zukunftskolleg Incoming Fellowship Programme, University of Konstanz, (Grant No. 291784), is gratefully acknowledged.

\appendix

\section{Classical dynamics and Langevin equation for the charge}
\label{app-A}

After a unitary transformation in Eq.~(\ref{eq:H_dual}), it is possible to show that
the charge $q_J$ on the QPSJ satisfies the Langevin equation
\begin{equation}
dq_J/dt = I_0 + I_\textrm{mw} (t) - I_{RL}(t) + \delta I(t) \, ,
\label{quantlang}
\end{equation}
as discussed in previous works, e.g. in Ref.~\onlinecite{Likharev:1985-1}.
Notice that $dq_J/dt=I_J$ corresponds to the current flowing through the QPSJ.
The first and second term in the right-hand side of Eq.~(\ref{quantlang}) give the total bias current.
The third term is the current flowing through the resistive-inductive branch of the circuit
$I_{RL}(t)=\int dt' Y(t-t') V_J(t')$, where $V_J(t) = V_c \sin\left[ \pi q_J(t) / e \right]$ and
$Y(t)$ is the inverse Fourier transform of the admittance Eq.~(\ref{adm}).
The last term of the right-hand side in Eq.~(\ref{quantlang}) is a fluctuating current $\delta I(t)$ of zero average whose Fourier component
satisfies the fluctuation-dissipation theorem
\begin{equation}
\langle \delta I(\omega)\delta I(\omega')\rangle = 2\pi \delta(\omega+\omega')\hbar \omega \Re \mbox{e}[Y(\omega)] \coth (\hbar \omega/2 k_B T),
\end{equation}
where $T$ is the temperature of the environment.
When $I_\textrm{mw} =0$, Eq.~(\ref{quantlang}) reduces to the well-known Langevin problem of the quasi-charge dynamics in the overdamped regime.
Then, the DC current-voltage characteristic of the QPSJ corresponds to the so-called Bloch nose (see Fig.~\ref{fig8}).

Disregarding, for instance, the fluctuation  $\delta I=0$ and considering the limit $g U_0/E_L \ll 1$ in Eq.~(\ref{quantlang}),
by direct integration, one obtains the following DC voltage
\begin{equation}
V_J^{(\delta I=0)}  = \frac{R_Q I_0}{g} - \theta\left( \frac{R_Q I_0}{g} - V_c \right) \sqrt{ {\left(  \frac{R_Q I_0}{g} \right)}^2 - V_c^2 }  \ ,
\label{classsol}
\end{equation}
where $\theta(V)$ is the Heaviside step function. The corresponding current-voltage characteristic is shown in the inset of Fig.~\ref{fig8}. It consists of a zero-current branch at finite voltage up to $V_c$ which bends back to a low-voltage, finite current branch.
Finite, classical charge fluctuations, $\delta I (t) \neq 0$, prevent the formation of a sharp feature in the current-voltage characteristic, even for small $g$,
and yielding a smearing of the Bloch nose.

\section{QPSJ in the absence of microwaves.}
\label{appb}

In this appendix, we consider the current-voltage characteristics of a QPSJ without microwave irradiation using Eqs.~(\ref{eq:Expansion}) and (\ref{eq:Expansion_TimeAverage}), and recall some of the results provided by the $P(E)-$theory,\cite{Ingold:1992} which describes the phase slippage in the presence of an external environment.~\cite{Averin:19902,Zazunov:2008} Setting $\alpha=0$ in Eq.~(\ref{eq:Expansion_TimeAverage}), and retaining the term $n=0$ only, the voltage drop on the QPSJ as a function of $\omega_B$ reads as
\begin{equation}
\label{eq:First_Order_NoBias}
\frac{V_J^{(\textrm{DC})}}{V_c} \left( \omega_B \right) \simeq \frac{\pi}{2}U_0\left[P(\hbar\omega_B) - P(-\hbar\omega_B) \right] \ ,
\end{equation}
where we defined the function\cite{Ingold:1992}
\begin{equation}
\label{eq:Dual_PE}
P\left( \Delta E \right)\equiv\frac{1}{2\pi\hbar}\int_{-\infty}^{+\infty}  d\tau \  e^{J(\tau)} \ e^{\frac{i}{\hbar } \Delta E \tau} \ .
\end{equation}
The function  $P(\Delta E)$ represents the probability density that the QPSJ absorbs ($\Delta E>0$) or emits ($\Delta E<0$) an amount of energy $|\Delta E|$ from or to the external environment respectively during a phase-slip event.
We see that an incoherent phase slippage by $\Delta \varphi=2\pi$ in the Wannier-Stark Ladder takes place only if the system exchanges the energy $\Delta E=\hbar\omega_B = (\Delta \varphi) \hbar I_0 /(2e)$ with the environment [see Fig.~\ref{fig3}(b)].
As the energy spectrum of the bath is continuous, the QPSJ has a dissipative behavior for any value of the applied DC current $I_0$.

%
%
\begin{figure}[hbtp]
{\includegraphics[scale=0.74]{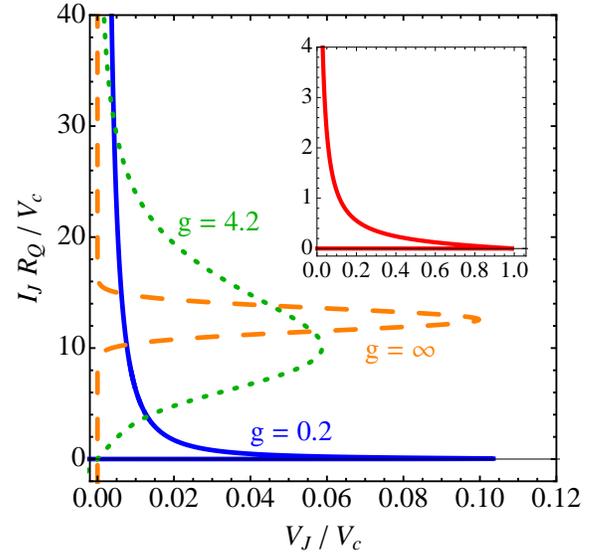}}
\caption{(Color online) Current-voltage characteristics of a QPSJ embedded in an inductive-resistive environment in the absence of microwaves.
Parameters: $k_BT/U_0 = 0.25$,  $U_0/E_L = 0.013$.
Curves from bottom to top correspond to $g=0.2$, 4.2, and $\infty$.
The inset shows the $I$-$V$ curve obtained by means of Eq.~(\ref{classsol}).
}
\label{fig8}
\end{figure}
%
%
%

We summarize the behavior of Eqs.~(\ref{eq:First_Order_NoBias}) and (\ref{eq:Dual_PE}) combined with Eqs.~(\ref{adm}) and (\ref{eq:J})
in the main panel of Fig.~\ref{fig8} where we plotted the QPSJ's current-voltage characteristic for various values of $g$ at low temperature
as obtained by direct numerical integration.

The validity of Eq.~(\ref{eq:First_Order_NoBias}) is given by the condition $V_J^{(\textrm{DC})}/V_c \ll 1$, yielding $U_0 \textrm{max}[P(\hbar \omega_B)]\ll 1$.~\cite{Ingold:1994}
Analytical results are available, for instance, in the limit of low temperature and small conductance so that $\beta E_L/ 2 \pi^2 g \gg 1$ and $\beta E_L/ 2 \pi^2 g^2 \gg 1$. Then,\cite{Ingold:1994,Zazunov:2008}
\begin{equation}
\frac{V_J^{(\textrm{DC})}}{V_c}(\omega_B) \simeq u \ \frac{|\Gamma(g+ i\beta \hbar \omega_B/2\pi)|^2}{\Gamma(2g)} \sinh (\beta \hbar \omega_B/2) \ ,
\end{equation}
where $u = (\beta U_0/4 \pi)(\beta E_L e^\gamma/2 \pi^2 g)^{-2 g}$ with $\gamma = 0.577\ldots$ the Euler constant. Hence we find a linear conductance $G_0$ at vanishing current $I_J$ and voltage $V_J$ ($\hbar\omega_B \beta /2\pi \ll 1$), given by
\begin{equation}
\label{eq:Slope_1}
G_0 R_Q \approx 4 (U_0 \beta)^{2(g-1)} \left(\frac{E_L}{U_0}\right)^{2 g} \left(\frac{1}{2\pi^2 g} \right)^{2g} \frac{\Gamma(2g)}{\Gamma^2(g)}e^{2g\gamma}-g  \  .
\end{equation}
We note that $G_0 \sim T^{2 -2g}$ and thus decreases with decreasing temperature; similarly $G_0 \sim E_L^{2 g} \sim 1/L^{2 g}$ and thus decreases with increasing inductance. Moreover, $G_0$ decreases with decreasing $g$.
%
%
Increasing $\omega_B$ until $\omega_B^\textrm{max} \approx  2\pi g / (\hbar \beta)$, we reach the back-bending point corresponding to the maximum value
\begin{equation}
\label{eq:Vj_max}
\left( \frac{V_J^{(\textrm{DC})}}{V_c} \right)_\textrm{max} \approx \pi \ u = \frac{1}{4}(\beta U_0)^{1 - 2g}\left(\frac{U_0}{E_L}\right)^{2 g} (2 \pi^2 g)^{2 g} e^{-2 \gamma g} \ ,
\end{equation}
for $g \ll 1$. We see that the lower the temperature $T$, the larger is the inductance $L$, and the smaller the conductance $g$, the closer $(V_J^{(\textrm{DC})}/V_c)_\textrm{max}$  is to the maximum value $V_c$. Beyond the back-bending point, corresponding to the maximum voltage, the system enters into the Bloch oscillation branch where the bias energy $\hbar \omega_B$ becomes dominant with respect to both quantum and thermal fluctuations and the DC voltage $V_J$ decreases exponentially to zero.

When the resistance $R$ is reduced so that $g>1$, the current-voltage characteristic is smeared into a smooth curve with a maximum voltage at finite current.
In the high-conductance regime $g\gg 1$,  we have that the Bloch nose broadens into a Gaussian:
\begin{equation}
\label{eq:gaussian}
P(\Delta E) \simeq \frac{1}{\sqrt{4 \pi E_L k_B T}} \exp\{-(\Delta E-E_L)^2/(4 E_L k_B T)\} \ .
\end{equation}
As a result, phase-slip events in a current-biased QPSJ can only occur if the energy $\hbar \pi I_0/e$ exchanged with the inductive environment equals $E_L$,
viz., the current $I_J$ at which the QPSJ sustains the largest voltage approaches the value $\Phi_0/2L$.
This phenomenon is dual to the Coulomb blockade found in a Josephson junction in a highly resistive environment,
where the voltage at which the Josephson junction sustains the largest current approaches the value $2e/2C$.~\cite{Zazunov:2008}

\section{QPSJ Hamiltonian for an underdamped Josephson junction}
\label{appa}

In this appendix, we will study the Hamiltonian of the circuit of Fig.~\ref{figapp1}, which is formed by a Josephson junction (JJ), biased by a time-dependent current $I(t) = I_0 + I_\textrm{mw} \cos \omega_\textrm{mw} t$, in parallel with a capacitance $C$ and an external electromagnetic environment composed by a resistance $R$ and an inductance $L$ in series. In particular, we will show that this Hamiltonian reduces to the QPSJ Hamiltonian (\ref{eq:H_dual}) used in the main text under suitable conditions.

Neglecting the contribution of the quasi-particle excitations, the Hamiltonian corresponding to the circuit of Fig.~\ref{figapp1} is given by the sum of the charging energy of the capacitance $C$, the non-linear Josephson energy and the energy of the environment,
\begin{equation}
\label{eq:H_Ibiasapp}
\hat{H}_s  = \frac{1}{2C}\left[ \int_{-\infty}^t dt' I(t')+\hat{Q}_\textrm{RL}+\hat{Q} \right]^2-E_J\cos\left( \hat{\varphi} \right)+\hat{H}_\textrm{env} \ .
\end{equation}
The phase operator $\hat{\varphi}$ is the phase-difference between the two superconductors forming the junction and $\hat{Q}$ is its conjugate charge operator $\left[ \hat{\varphi},  \hat{Q} \right] = 2e \, i$, i.e., the charge tunneling through the junction. In Eq.~(\ref{eq:H_Ibiasapp}), we also introduced $\hat{Q}_\textrm{RL}=\sum_\lambda \hat{Q}_\lambda$ which accounts for the charge noise produced by the $R$-$L$ environment, as discussed in the main text.

%
%
\begin{figure}[hbtp]
\includegraphics[scale=0.5,angle=0.]{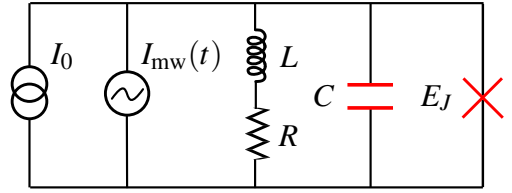}
\caption{Current-biased Josephson junction with Josephson energy $E_J$ in parallel with a capacitance $C$ and embedded in a resistive ($R$) and inductive ($L$) electromagnetic environment. The circuit is biased with a time-dependent current $I(t)$.}
\label{figapp1}
\end{figure}
%
%
%

The equivalence between Hamiltonian (\ref{eq:H_Ibiasapp}) and QPSJ-Hamiltonian (\ref{eq:H_dual}) can be demonstrated through the following steps. First, we apply the gauge and the unitary transformations $\hat{U}_g(t)=\exp\left[  -i\hat{\varphi}\int^t dt' I(t')/2e  \right]$ and $\hat{U}_\textrm{env}=\exp\left[- i\hat{\varphi}\hat{Q}_\textrm{RL}/2e  \right]$, respectively, to Eq.~(\ref{eq:H_Ibiasapp}) and we get
\begin{equation}
\label{eq:H_Ibias_1app}
\hat{H}^{'}_s= \frac{\hat{Q}^2}{2C}-E_J\cos(\hat{\varphi})  - \frac{\hbar I(t)}{2e}\hat{\varphi} +
\hat{H}_\textrm{env} \left[ \{\hat{Q}_{\lambda}\}, \{ {\hat{\varphi}_{\lambda} + \hat{\varphi}} \}  \right] \, .
\end{equation}
Here, the first and second term correspond to the standard Hamiltonian $\hat{H}_J$ of an isolated JJ. In the tight-binding regime, $E_J \gg E_C$, $\hat{H}^{'}_s$ becomes
\begin{equation}
\label{eq:H_Ibias_2app}
\hat{H}^{''}_s = -U_0 \cos\left( \frac{\pi}{e} \hat{q} \right) - \frac{\hbar I(t)}{2e}\hat{\varphi} +
\hat{H}_\textrm{env} \left[ \{\hat{Q}_{\lambda}\}, \{ {\hat{\varphi}_{\lambda} + \hat{\varphi}} \}  \right] \, .
\end{equation}
where  $\hat{q}$ is the quasi-charge operator and $U_0=8\sqrt{E_J \hbar\omega_p/\pi}\exp{(-\sqrt{8E_J/E_C})}=e V_c / \pi$ the half-bandwidth of the first Bloch band of $\hat{H}_J$. Within this limit, an energy gap of the order of the plasma frequency $\hbar\omega_p=\sqrt{8E_J E_C}$ separates the first from the second Bloch band. We neglect the possibility of inter-band Landau-Zener transitions assuming the low temperature and bias current limit $(k_BT, \hbar I_0/2e,  \hbar I_{\textrm{mw}}/2e)  \ll  \hbar\omega_p$ as well as considering an off-resonance microwave field, $\omega_{\textrm{mw}} \ll \omega_p$.

Finally, we apply the inverse unitary transformation $\hat{U}_\textrm{env}^{-1}$ to Eq.~(\ref{eq:H_Ibias_2app}) and we obtain the effective low-energy Hamiltonian
\begin{equation}
\label{eq:H_dualapp}
\hat{H} \! = \! -U_0 \cos\left[ \frac{\pi}{e}  \left( \hat{q}  + \hat{Q}_{RL} \right) \right]
- \frac{\hbar I(t)}{2e}\hat{\varphi} +
\hat{H}_\textrm{env} \left[ \{\hat{Q}_{\lambda}\}, \{ \hat{\varphi}_{\lambda} \}  \right]  \ .
\end{equation}
This is the energy operator (\ref{eq:H_dual}) of the main text describing a current-biased quantum phase-slip junction coupled to an external $R$-$L$ electromagnetic environment, as depicted in Fig.~\ref{fig1}(a).

\bibliography{references}

\end{document}